\documentclass[prl,twocolumn]{revtex4-1} 
\usepackage[utf8]{inputenc}
\usepackage{graphicx} 
\usepackage{amsmath} 
\usepackage{amssymb} 
\usepackage{multirow}
\usepackage{ulem}
\usepackage{hyperref}
\hypersetup{
  pdfauthor={Harald G. L. Schwefel},
  pdftitle={Efficient single sideband microwave to optical conversion using an electro-optical whispering gallery mode resonator},
  pdfsubject={Physics, Optics, Whispering Gallery Mode Resonators, electro-optical modulator, nonlinear optics, microwave, coherent frequency conversion},
  urlcolor=blue,
}

\usepackage{siunitx}
\usepackage{color}
\newcommand{\unit}[1]{\,#1}

\newcommand{\nii}[1]{_{\mathrm{#1}}}
\newcommand{\authoro}[2][]{\author{#2$^{#1}$}}    
\newcommand{\affil}[2][]{\affiliation{$^{#1}$#2}} 
\begin{document} 
\title{Efficient single sideband microwave to optical conversion using an electro-optical whispering gallery mode resonator}

 \authoro[1,2,3,+]{Alfredo Rueda}
 \authoro[1,2,3,+,*]{Florian Sedlmeir}
 \authoro[1,2,4,5]{Michele C.\ Collodo}
 \authoro[1,2]{Ulrich Vogl}
 \authoro[1,2,6]{Birgit Stiller}
 \authoro[1,2,3]{Gerhard Schunk}
 \authoro[1]{Dmitry V.\ Strekalov}
 \authoro[1,2]{Christoph Marquardt}
 \authoro[4,7]{Johannes M.\ Fink}
 \authoro[4]{Oskar Painter}
 \authoro[1,2]{Gerd Leuchs}
 \authoro[8,*]{Harald G.\ L.\ Schwefel\vspace{2ex}}
 \affil[1]{Max Planck Institute for the Science of Light, Günther-Scharowsky-Straße 1/Building 24, 90158 Erlangen, Germany}
\affil[2]{Institute for Optics, Information and Photonics, University Erlangen-Nürnberg, Staudtstr.\ 7/B2, 91058 Erlangen, Germany}
\affil[3]{SAOT, School in Advanced Optical Technologies, Paul-Gordan-Str.\ 6, 91052 Erlangen, Germany}
\affil[4]{Institute for Quantum Information and Matter and Thomas J. Watson, Sr., Laboratory of
Applied Physics, California Institute of Technology, Pasadena, California 91125, USA}
\affil[5]{currently at: Department of Physics, ETH Zürich, CH-8093 Zurich, Switzerland}
\affil[6]{currently at: Centre for Ultrahigh bandwidth Devices for Optical Systems (CUDOS), School of Physics, University of Sydney, New South Wales 2006, Australia}
\affil[7]{currently at: Institute of Science and Technology Austria, 3400 Klosterneuburg, Austria}
\affil[8]{Department of Physics, University of Otago, Dunedin, New Zealand\vspace{1ex}}

\affil[*]{Corresponding authors: Florian.Sedlmeir@mpl.mpg.de, Harald.Schwefel@otago.ac.nz}

\affil[+]{these authors contributed equally to this work}

\begin{abstract}
Linking classical microwave electrical circuits to the optical telecommunication band is at the core of modern communication. Future quantum information networks will require coherent microwave-to-optical conversion to link electronic quantum processors and memories via low-loss optical telecommunication networks. Efficient conversion can be achieved with electro-optical modulators operating at the single microwave photon level. In the standard electro-optic modulation scheme this is impossible because both, up- and downconverted, sidebands are necessarily present. 
Here we demonstrate true single sideband up- or downconversion in a triply resonant whispering gallery mode resonator by explicitly addressing modes with asymmetric free spectral range.
Compared to previous experiments, we show a three orders of magnitude improvement of the electro-optical conversion efficiency reaching 0.1\% photon number conversion for a 10\,GHz microwave tone at 0.42\unit{mW} of optical pump power. The presented scheme is fully compatible with existing superconducting 3D circuit quantum electrodynamics technology and can be used for non-classical state conversion and communication. Our conversion bandwidth is larger than 1\,MHz and not fundamentally limited.
\end{abstract}

\maketitle

\section{Introduction}
Efficient conversion of signals between the microwave and the optical domain is a key feature required in classical and quantum communication networks.
In classical communication technology, the information is processed electronically at microwave frequencies of several gigahertz. Due to high ohmic losses at such frequencies, the distribution of the computational output over long distances is usually performed at optical frequencies in glass fibers.
The emerging quantum information technology has developed similar requirements with additional layers of complexity, concerned with preventing loss, decoherence, and dephasing.
Superconducting qubits operating at gigahertz frequencies are promising candidates for scalable quantum processors \cite{schoelkopf_wiring_2008,devoret_superconducting_2013,riste_detecting_2015}, while the optical domain offers access to a large set of very well developed quantum optical tools, such as highly efficient single-photon detectors and long-lived quantum memories \cite{lvovsky_optical_2009}. Optical communication channels allow for low transmission losses and are, in contrast to electronic ones, not thermally occupied at room temperature due to the high photon energies compared to $k_{\text B} T$.\\
To be useful for quantum information, a noiseless conversion channel with unity conversion probability has to be developed. For the required coupling between the vastly different wavelengths, different experimental platforms have been proposed, e.g. cold atoms \cite{hafezi_atomic_2012,verdu_strong_2009}, spin ensembles coupled to superconducting circuits \cite{imamoglu_cavity_2009,marcos_coupling_2010}, and trapped ions \cite{williamson_magneto-optic_2014,fernandez-gonzalvo_frequency_2015}. The highest conversion efficiency so far was reached via electro-optomechanical coupling, where a high-quality mechanical membrane provides the link between an electronic LC circuit and laser light \cite{bagci_optical_2014,andrews_bidirectional_2014}. Nearly 10\% photon number conversion from \SI{7}{GHz} microwaves into the optical domain was demonstrated in a cryogenic environment \cite{andrews_bidirectional_2014}. Electro-optic modulation as a different approach has been recently discussed by Tsang \cite{tsang_cavity_2010,tsang_cavity_2011}. In this scheme the refractive index of a transparent dielectric is modulated by a microwave field. 
The resulting phase modulation creates sidebands symmetrically around the optical pump frequency. In the photon picture this can be understood as coherent superposition of two processes. The process of sum-frequency generation (SFG) combines a microwave and an optical photon and creates a blue-shifted photon (anti-Stokes). This process is fundamentally noiseless and hence does not lead to decoherence if the photon conversion efficiency approaches unity.\\
In the process of difference frequency generation (DFG) the microwave signal photon can stimulate an optical pump photon to decay into a red-shifted optical (Stokes) and an additional microwave photon. Such microwave parametric amplification adds a minimum amount of noise as the process can also occur spontaneously (spontaneous parametric down-conversion) without the presence of a microwave signal \cite{caves_quantum_1982}. 
This spontaneous process can be used to generate entangled pairs of microwave and optical photons \cite{tsang_cavity_2011}. Above threshold the spontaneous process can stimulate parametric oscillations generating coherent microwave radiation \cite{savchenkov_parametric_2007}.\\
Significant improvements of electro-optic modulation were made in the recent years by developing resonant modulators with high quality ($Q$) factors consisting of a lithium niobate whispering gallery mode resonator (WGM) coupled to a microwave resonator \cite{cohen_microphotonic_2001-1,ilchenko_whispering-gallery-mode_2003}. The best reported photon conversion efficiency so far was of the order of 0.0001\% per mW optical pump power \cite{strekalov_efficient_2009,ilchenko_whispering-gallery-mode_2003}. These implementations could not individually address either SFG or DFG and therefore always include additional noise from the spontaneous process.\\
Electro-optic single-sideband conversion has been demonstrated in a lithium tantalate WGM resonator, where the optical pump and the optical signal were orthogonally polarized (type-I conversion) \cite{savchenkov_tunable_2009}. The highest efficiency in lithium niobate can be expected when signal and pump are parallelly polarized (type-0 conversion). It was discussed that single sideband conversion in this case can be achieved by detuning the optical pump from its resonance \cite{tsang_cavity_2010} (see Fig.~\ref{fig:assymetric_fsr_scheme}(a)), which increases the required pump power significantly.\\
We present a new take on the classical electro-optical modulation within a lithium niobate whispering gallery mode resonator, where single sideband operation is made possible without detuning. The demonstrated photon number conversion efficiency of 0.2\% per milliwatt optical pump power is three orders of magnitude larger than in previous works \cite{strekalov_efficient_2009,ilchenko_whispering-gallery-mode_2003}. {The significant increase of the conversion efficiency is due to better optical and microwave $Q$ factors and enhanced model overlap. Embedding the WGM resonator within a closed three dimensional microwave cavity allows to strongly focus the microwave field into the optical mode volume. Furthermore, our system is  fully compatible with  superconducting circuit QED \cite{rigetti_superconducting_2012,paik_observation_2011}.}\\
We are able to switch between sum- and difference frequency generation by employing avoided crossings of the optical modes, which lead to an asymmetric spectral distance between three neighboring resonances. {Since the mode crossings are temperature dependent, they can be used to tune the microwave frequency of the converter over several tens of MHz.\\
Current state of the art electro-optomechanical conversion schemes have a bandwidth of several kHz \cite{andrews_bidirectional_2014,bagci_optical_2014} and their bandwidth is fundamentally limited to the mechanical resonance frequency. The electro-optical scheme does not have this fundamental limitation. In the current implementation we already reach a bandwidth of 1\,MHz.}   

\section{All-resonant electro-optics}
\label{sec:theory}
\begin{figure}
\centering
\includegraphics[clip,angle=0,width=0.9\linewidth]{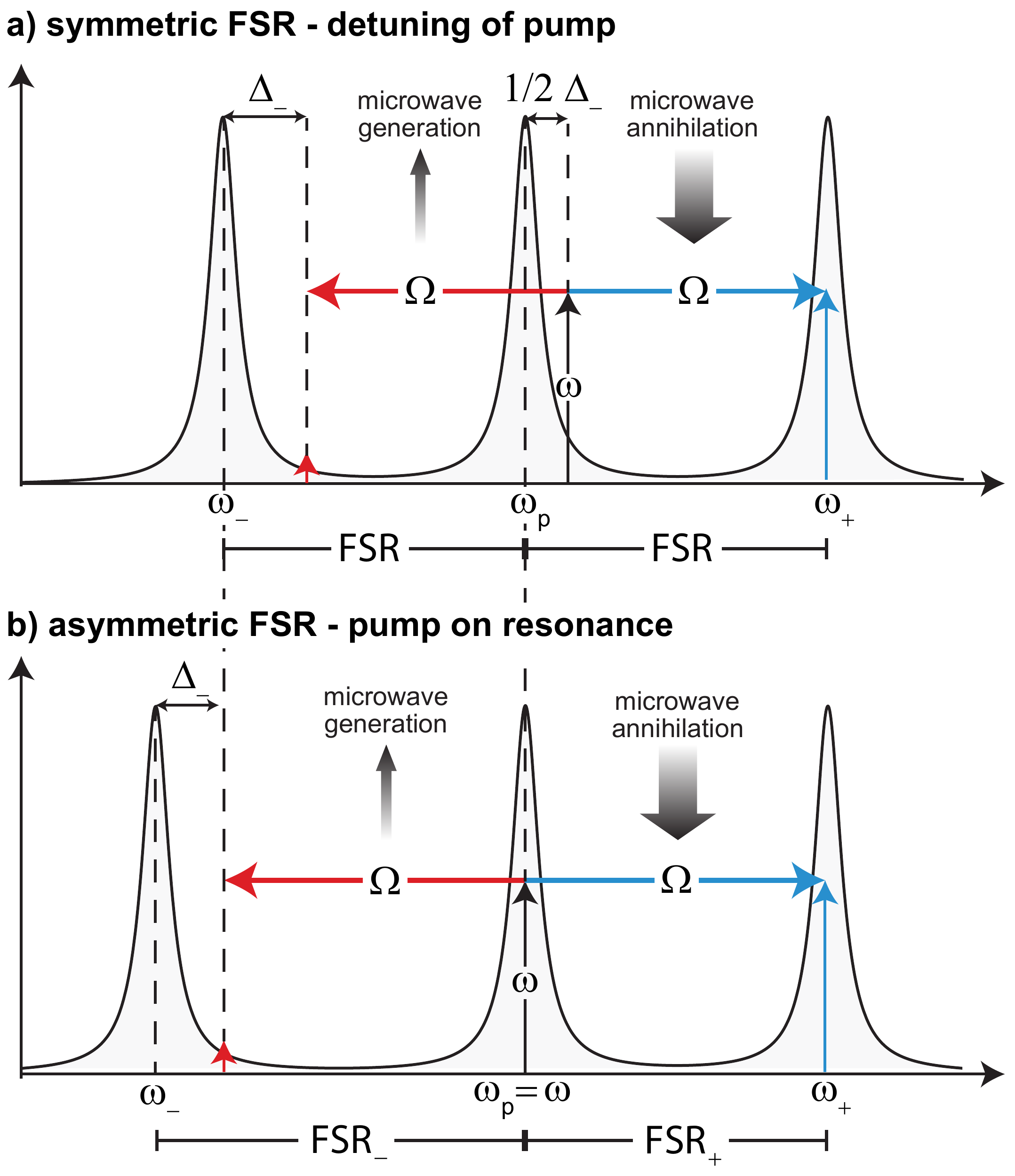}
\caption[Modes]{Two schemes for suppressing the DFG signal in type-0 conversion are shown. In the case of equally spaced modes (a), the optical pump has to be detuned to cause the red sideband to be off resonant. If the mode spacing is asymmetric (b), the pump can be kept fully resonant while maintaining the suppression of the undesired sideband.}
\label{fig:assymetric_fsr_scheme}%
\end{figure}
Our conversion scheme is based on the nonlinear interaction between resonantly enhanced microwave and optical fields within a high quality WGM resonator. The optical resonator is made out of lithium niobate and placed within a three dimensional copper cavity. While the optical mode is guided at the rim of the WGM resonator by total internal reflection, the microwave field is confined by the metallic boundary of the copper cavity which is designed to enforce good overlap of the optical and the microwave modes within the lithium niobate. As lithium niobate is an electro-optic material, the microwave field modulates the refractive index and couples parametrically to the optical modes. Assuming weak microwave fields, maximally two sidebands, the up- and the downconverted one, will be generated next to the optical pump frequency. The interaction Hamiltonian describing the system is
\begin{figure*}[t!]
	\centering
	\includegraphics[width=1\linewidth]{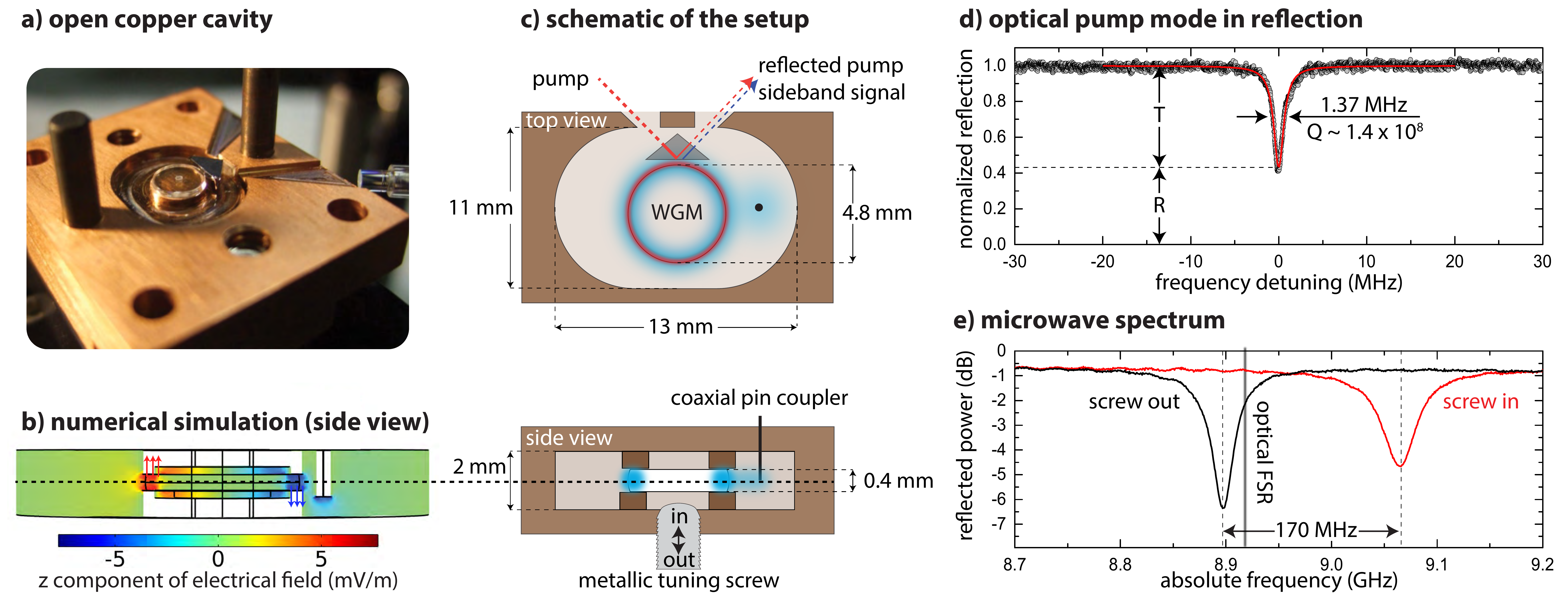}
	\caption{a) A photograph of the bottom part of the microwave cavity with silicon coupling prism and the WGM resonator. b) Simulation of the microwave field distribution in the cavity. Only the $z$-component (TE) of the field is shown. c) {Schematic of the cavity. An optical pump beam (red dashed) couples through a prism into the WGM and part of the light is directly reflected. The sideband is given in blue.} In the side view, the metallic tuning screw to perturb the microwave mode and the coaxial pin coupler are shown. The pin coupler is defined as the input port of the converter, the output port is the optical outcoupling spot inside of the prism (solid lines). d) Reflection spectrum of the optical mode used for sum-frequency conversion (compare Fig.~\ref{fig:OSA}). Also shown are the linewidth and the corresponding $Q$ factor. e) Microwave spectrum in reflection from the coaxial pin coupler. The $m_\Omega = \pm 1$ mode is shown for tuning screw position inside (red) and outside (black) the cavity.}
	\label{fig:setup}
\end{figure*}\begin{equation}
\hat{H}_\text{int} =  \hbar g (\underbrace{\hat{a} \hat{b}^\dagger_- \hat{c}^\dagger +\hat{a}^\dagger \hat{b}_- \hat{c}}_{\text{DFG}} + \underbrace{\hat{a} \hat{b}^\dagger_+ \hat{c} +\hat{a}^\dagger \hat{b}_+ \hat{c}^\dagger}_{\text{SFG}}),
\label{equ:interHamiltonian}%
\end{equation}
where the operators $\hat{a},\, \hat{b}_+,\, \hat{b}_-$ denote the optical pump, the up- and downconverted optical sidebands, respectively, and $\hat{c}$ the microwave mode. The coupling rate $g_\pm$ is given by (compare e.g. \cite{ilchenko_whispering-gallery-mode_2003})
\begin{equation}
g_\pm =  \frac{n\nii{p}n_\pm}{n_\Omega} r \sqrt{\frac{\hbar\omega\nii{p}\omega_\pm \Omega_0}{8 \epsilon_0 V\nii{p}V_\pm V_\Omega}} \times \int \!\text{d}V \,\,\,  \Psi\nii{p}\Psi_\Omega \Psi_\pm,
\label{equ:g}
\end{equation}
where $n\nii{p}$, $n_\pm$, and $n_\Omega$ are the refractive indices of pump, sideband, and  microwave tone and $\omega\nii{p}$, $\omega_\pm$, and $\Omega_0$ are the respective resonance frequencies. The normalization volumes $V\nii{p}$, $V_\pm$, and $V_\Omega$ are given by the integral $\int \text{d}V \Psi \Psi^*$ over the respective spatial field distributions $\Psi\nii{p}$, $\Psi_\pm$, and $\Psi_\Omega$. The nonlinear coupling is mediated by the electro-optic coefficient $r$. The integral in \eqref{equ:g} is non-zero only if phase-matching is fulfilled. In the case of whispering gallery modes this means that the azimuthal mode numbers $m$ (number of wavelengths around the rim of the resonator) have to obey the relation $m_\pm = m\nii{p}\pm m_\Omega$. In our experiment, the optical pump $P\nii{p}$ and the microwave signal $P_\Omega$ are both undepleted {(i.e.\ the nonlinear conversion efficiency depends linearly on the pump and signal intensities)}. From \eqref{equ:interHamiltonian} we find the output power of the sidebands $P_\pm$ and the photon-number conversion efficiency $\eta_\pm$ to be:
\begin{equation}
P_\pm = \underbrace{\frac{8 g^2}{\hbar \Omega} \frac{\gamma^2 \gamma_\Omega}{|\Gamma\nii{p}|^2 |\Gamma_\pm|^2 |\Gamma_\Omega|^2}}_{\zeta_\pm} P\nii{p}P_\Omega \quad \text{and} \quad \eta_\pm = \frac{\Omega}{\omega_\pm} \zeta_\pm  P\nii{p},
\label{power_conversion}
\end{equation}
where
\begin{subequations}
\begin{alignat}{2}
\Gamma\nii{p}=& - \mathrm{i} (\omega - \omega\nii{p}) + \gamma + \gamma' \\
\Gamma_\Omega =& - \mathrm{i} (\Omega - \Omega_0) + \gamma_\Omega + \gamma'_\Omega.\\
\Gamma_\pm =& - \mathrm{i} (\omega \pm \Omega - \omega_\pm) + \gamma + \gamma'
\end{alignat}
\end{subequations}
Expressions (4) describe the detuning of the optical pump $\omega$, the microwave $\Omega$ and the generated sidebands $\omega \pm \Omega$ from their respective resonance frequencies. {Since the optical modes have the same polarization and are close in frequency, we can assume the same coupling and loss rates $\gamma$ and $\gamma'$ for both of them. Coupling and loss rate for the microwave mode are denoted as $\gamma_\Omega$ and $\gamma'_\Omega$.}\\
If the optical modes are spectrally equidistant, the suppression of one of the sidebands can be achieved by detuning the pump from its resonance frequency as schematically depicted in Fig.~\ref{fig:assymetric_fsr_scheme}(a). However, detuning decreases the coupling of the pump to the cavity and therefore increases the required pump power significantly (see below). We propose an alternative scheme depicted in Fig.~\ref{fig:assymetric_fsr_scheme}(b): the spectral distance, or free spectral range (FSR), can be asymmetric for the up- and downconverted modes. This would allow detuning of, for example the downconverted light, by $\Delta_- = \omega - \Omega - \omega_-$, while the pump and the up-converted light remain fully resonant. We will show, that such dispersion engineering is possible by exploiting avoided crossings. Both schemes in Fig.~\ref{fig:assymetric_fsr_scheme} lead to a power suppression factor $S$ between the sidebands of  
\begin{equation}
S = \frac{P_+}{P_-} = \frac{\Delta_-^2}{(\gamma + \gamma')^2}+1.
\label{equ:suppression}
\end{equation}
As the suppression depends only on the detuning of the downconverted sideband from its resonance, the two schemes can be compared in terms of the required pump power using \eqref{power_conversion}. {One can show with the help of Eqs.~(\ref{power_conversion}) and (\ref{equ:suppression}) that the detuning scheme requires $(S - 1)/4 + 1$ times more pump power to produce the same suppression $S$. For example, \SI{30}{dB} suppression would require about 250 times more pump power.}

\section{Experimental realization}
The challenge in this experiment is to bring two resonant systems having largely different frequencies to interact. The design has to provide a good spatial overlap between microwave and optical modes while the involved modes have to fulfill phase-matching and energy conservation. This requires careful tailoring of the system based on numerical simulations.\\
For the experiment, we fabricated a lithium niobate WGM resonator with a radius of $R = \SI{2.4}{mm}$ and a thickness $d = \SI{0.4}{mm}$ by single point diamond turning and subsequent polishing \cite{grudinin_ultra_2006}. The resonator has a z-cut configuration where the optic axis is aligned parallel to the symmetry axis. The disk is placed in a 3D microwave copper cavity as depicted in Fig.~\ref{fig:setup}(a-c) where it is clamped by two metallic rings that are designed to maximize the field overlap between the microwave and the optical modes. A silicon prism is placed within the cavity touching the WGM resonator for optical coupling. To enable critical coupling, the coupling plane of the prism is coated with a thermally grown silicon oxide layer serving as a spacer. As optical pump we use a narrow-band tunable laser ($\lambda \approx \SI{1550}{nm}$) which can either be swept over or locked to a resonance via the Pound-Drever-Hall technique. Two holes in the copper cavity allow the optical pump light to enter and the light reflected and emitted from the WGM resonator to leave the cavity. This light is collected and analyzed by a photodiode and an optical spectrum analyzer (OSA).
The microwave signal is coupled via a coaxial pin coupler mounted close to the WGM resonator into the cavity and analyzed with a vector network analyzer (VNA) in reflection. A metallic tuning screw is used to perturb the microwave field for fine adjustment of its resonance frequency. The whole setup is thermally stabilized at room temperature with mK precision by a proportional-integral-derivative (PID) controller and a thermoelectric element attached on the outer side of the closed copper cavity.\\
All fields are primarily polarized along the optic axis of the WGM resonator (TE type modes) as we aim for type-0 conversion. This allows us to address the largest electro-optic tensor element of lithium niobate ($r_{33} \approx \SI{31}{pm/V}$ \cite{wong2002properties}).\\
Fig.~\ref{fig:setup}(d) shows the tuning over a typical optical mode detected in reflection. We find coupling efficiencies of {about 60\%. Although the modes are critically coupled ($\gamma = \gamma'$), 40\% of the light is reflected from the cavity due to imperfect spatial mode matching.}
Typically, {the loaded} $Q$ is larger than $1\times10^8$ corresponding to linewidths less than \SI{2}{MHz}. The optical free spectral ranges, which were measured with sideband spectroscopy \cite{Li:12}, are around \SI{8.95}{GHz}. \\
With the VNA the phase and amplitude of the microwave signal back-reflected from the cavity was measured. Fig.~\ref{fig:setup}(e) shows that the tuning screw allows to change the microwave resonance frequency from $\Omega_0 = 2 \pi \times \SI{8.90}{GHz}$ to $2 \pi \times\SI{9.07}{GHz}$. We find a loaded $Q = 246$ for the undisturbed mode, which decreases to $Q = 174$ for maximum perturbation by the tuning screw. Measurements of the phase response indicate that the microwave mode is {undercoupled ($\gamma_\Omega < \gamma'_\Omega$)} resulting in a non-ideal coupling efficiency around $60 - 75\%$, depending on the position of the tuning screw.\\
Note, that the pin coupler in our current setup excites a propagating ($m_\Omega = 1$) and a counter propagating ($m_\Omega = -1$) microwave mode and only the part propagating in the direction of the excited optical modes is converted. Considering these geometries and the analytic knowledge of the optical modes \cite{Breunig:13} we can rewrite \eqref{equ:g} as $g_\pm =n\nii{p}^2 \omega\nii{p} r_{33}E_\Omega({\bf r})/2$, where the single photon electric field $E_\Omega({\bf r})$ is determined from the simulated microwave field distribution in the region of the optical mode {(compare Fig.~\ref{fig:setup}(b)).} From this we estimate the {single photon coupling rate} for the $m = 1$ mode to be $g \approx 2 \pi \times \SI{28}{Hz}$.

\subsection*{Asymmetric FSR around avoided crossings }
\label{sec:avoided}
\begin{figure}
	\centering
	\includegraphics[width=0.96\linewidth]{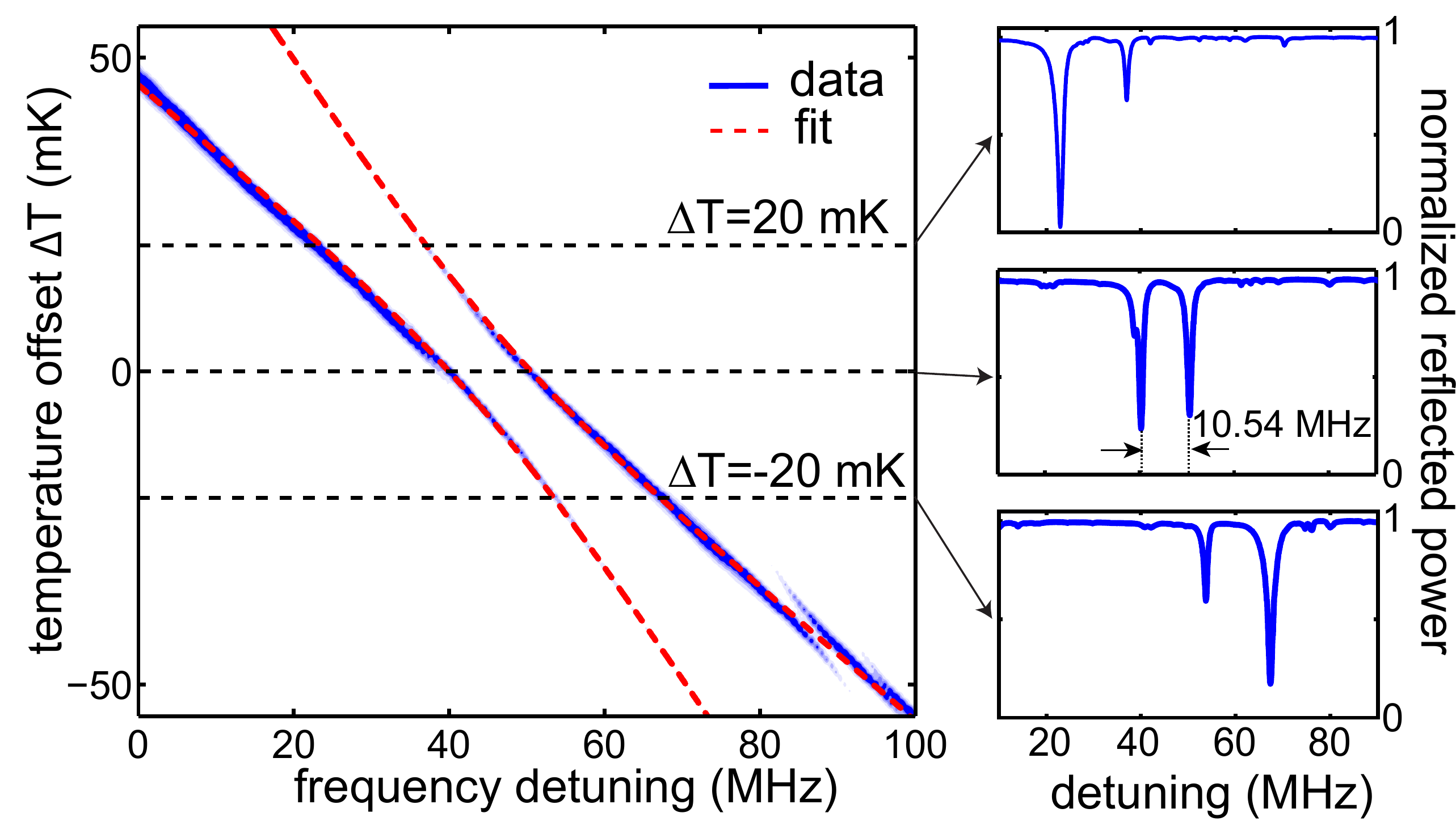}
	\caption{On the left, the resonance position of a TE mode experiencing an avoided crossing with a TM mode is shown for different temperature settings. On the right, the WGM spectrum around the avoided crossing is plotted for different temperatures.}
	\label{fig:av_crossing}
\end{figure}
\begin{figure*}[t]
	\centering
	\includegraphics[width=1\linewidth]{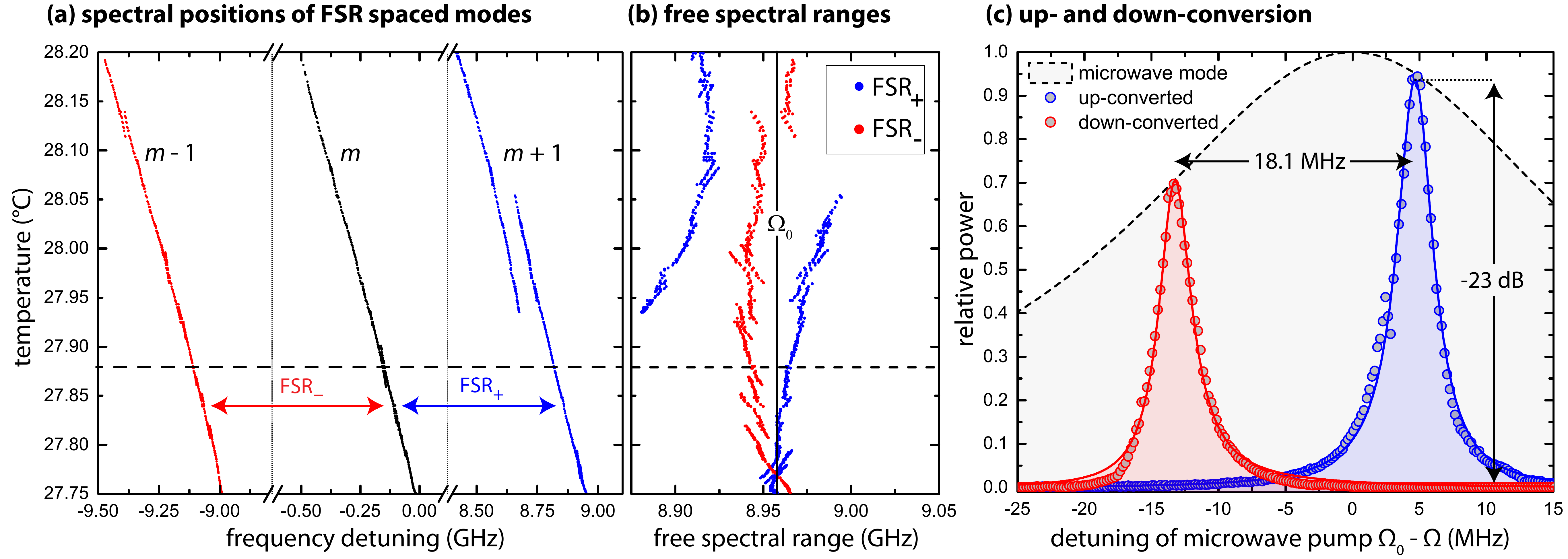}
	\caption{(a) Resonance frequencies of three TE modes that are separated by one azimuthal mode number $m$ corresponding to the free spectral range $\text{FSR}_\pm \approx \SI{9}{GHz}$. Each mode experiences avoided crossings of different strength for some temperatures. (b) Frequency difference of the $m+1$ and $m-1$ mode from the $m$ mode. One can see that $\textrm{FSR}_+$ and $\textrm{FSR}_-$ are functions of the temperature and differ by up to \SI{50}{MHz}. This allows selective up- and downconversion as depicted in (c): The temperature was set to \SI{27.88}{\degreeCelsius}, indicated by the dashed line in (a) and (b), where $\text{FSR}_+ - \text{FSR}_- = \SI{18.1}{MHz}$. The optical pump is locked to the $m$ mode. A microwave signal sweeping over the microwave resonance which is indicated by the gray Lorentzian, while the generated sidebands are measured with an optical spectrum analyzer (see Fig.~\ref{fig:OSA}). The shown sidebands are separated by \SI{18.1}{MHz} and the suppression of the down-conversion at the maximum of the up-conversion is about \SI{23}{dB}.}
	\label{fig:overview}
\end{figure*}
The key feature required for the single sideband conversion in a WGM resonator with type-0 phase-matching without detuning the pump is an asymmetric spectral distance between the pump and the two signal modes (compare Fig.~\ref{fig:assymetric_fsr_scheme}). In a mm-size resonator adjacent free spectral ranges differ only by a couple of kHz due to the combined effect of material and geometric dispersion. Hence, both sidebands are generated in the case of zero pump detuning. However, in WGM resonators different modes can couple to each other linearly if they are close to degeneracy \cite{carmon_static_2008,PhysRevLett.97.253901}. Because of different thermal dispersion of different WGM families, such modes can be brought into resonance by changing the resonator temperature. This feature allows them to exchange energy and leads to avoided crossings. Such crossings appear as a mode splitting around the unperturbed resonance frequency, where the minimal spectral separation is proportional to the coupling rate between the two modes and can reach many linewidths. Essentially, the modes shift away from their unperturbed resonance position leading to a change of the effective dispersion.\\
Such a crossing is shown in Fig.~\ref{fig:av_crossing} where the frequency of a TE polarized mode within a sweeping window of \SI{100}{MHz} is plotted against a temperature change of the resonator. From the fit in Fig.~\ref{fig:av_crossing} we extract a coupling rate of $\kappa = 2 \pi \times \SI{5.27}{MHz}$ and the asymptotes yielding the temperature dependence of the two interacting modes. The ratio of their slopes is approximately $1.8$, which corresponds to the ratio between the temperature sensitivities of TE and TM polarized modes in a lithium niobate WGM resonator of that size \cite{PhysRevB.50.751}. Such polarization coupling has been previously observed in ring resonators \cite{ramelow_strong_2014} and in larger magnesium fluoride resonators \cite{weng_mode-interactions_2015}.\\
To demonstrate the effect of the splitting on the spectral distance between two adjacent modes of the same family we show a spectrum of the $m-1,\, m$, and $m+1$ modes in Fig.~\ref{fig:overview}(a). The corresponding spectral distances $\text{FSR}_+ = \nu_{m+1}-\nu_m$ and $\text{FSR}_- = \nu_{m} - \nu_{m-1}$ are presented in Fig.~\ref{fig:overview}(b). The free spectral ranges are strongly temperature-dependent close to the avoided crossings and the difference $|\Delta_+ - \Delta_-|$ can be several tens of MHz, exceeding the linewidth of the modes significantly. We will use this temperature-dependent asymmetry of the FSR to switch between sum- and difference frequency generation in the following.

\subsection*{Single sideband conversion}
For conversion of a microwave signal to a single optical sideband we use the mode triplet shown in Fig.~\ref{fig:overview}(a) and (b). The optical pump laser was locked to the central mode denoted with $m$ (black curve) while the temperature of the cavity was stabilized around \SI{27.88}{\celsius}. The free spectral ranges at that working point are marked by the dashed line in Fig.~\ref{fig:overview}(b). The microwave cavity resonance was set to $\Omega_0 = 2\pi\times8.960$ GHz and the microwave signal sent to the cavity was swept from \SI{8.925}{GHz} to \SI{8.975}{GHz} in steps of \SI{200}{kHz}. For each step the optical signal was measured with an optical spectrum analyzer (YOKOGAWA AQ6370C) (see inset in Fig.~\ref{fig:OSA}). The obtained magnitude of the generated sidebands is shown in Fig.~\ref{fig:overview}(c) as a function of the microwave detuning from its resonance, which is indicated by the black dashed line.\\
The up- and downconverted signals can be addressed separately as they appear at different microwave frequencies due to the different FSRs of the respective modes. The Lorentzian given by \eqref{power_conversion} agrees well with the measured data. The asymmetry of the peaks and their intensity difference can be attributed to the detuning from the microwave resonance position. 
The central frequencies of the peaks are separated by \SI{18.1}{MHz} which corresponds to the difference between $\text{FSR}_+$ and $\text{FSR}_-$.  
The extinction ratio for the suppressed sideband is determined by the FSR-asymmetry and the linewidth of the microwave and optical modes. From the fit we find a suppression of the down-converted signal at maximum up-conversion of $23$\unit{dB}. This suppression could be further increased by a stronger mode splitting and narrower bandwidths of both, the microwave and the optical modes.

\section{Conversion efficiency}
\begin{figure}[h!!]
	\centering
		\includegraphics[width=1.0\linewidth]{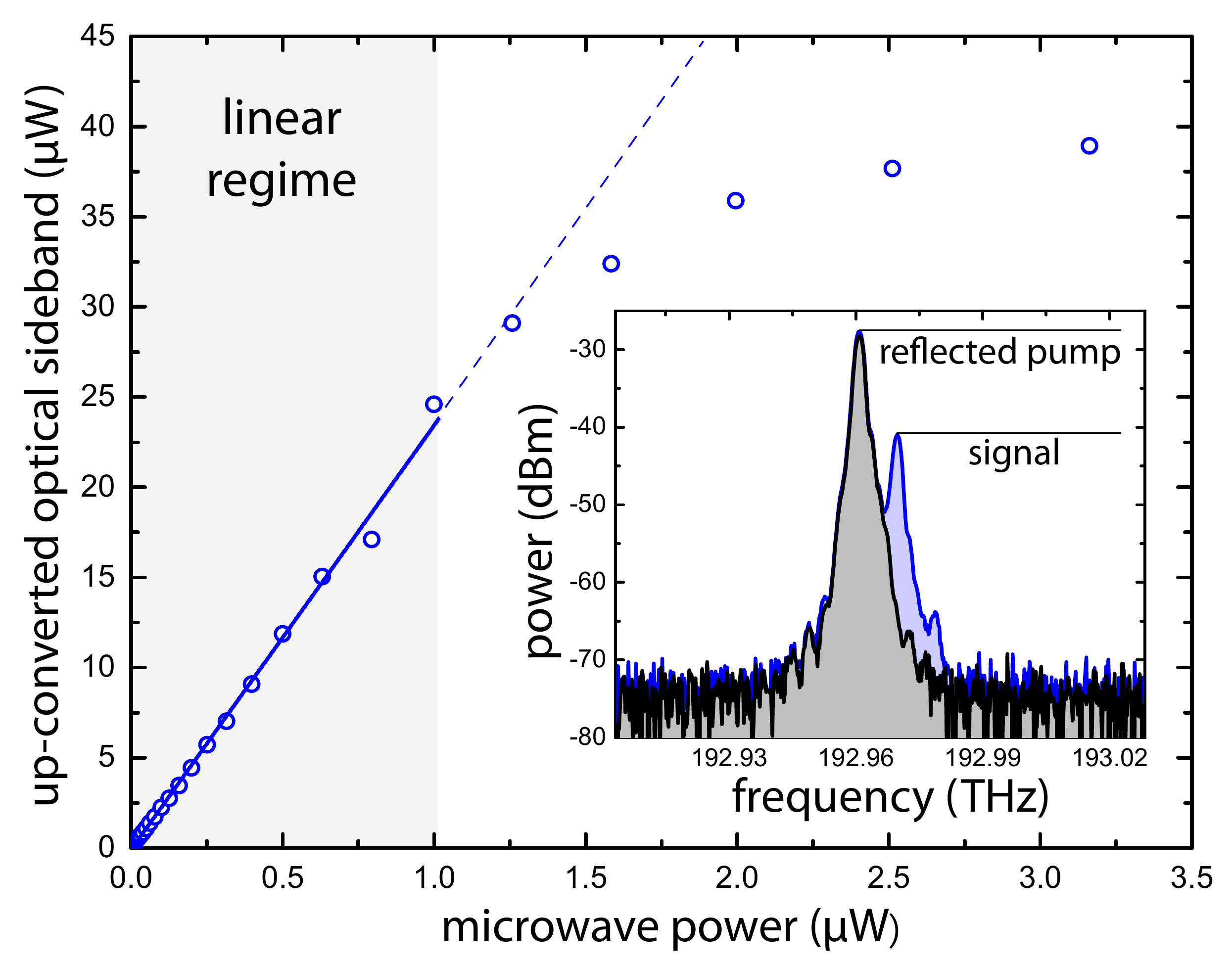}
	\caption{Power of the up-converted sideband as a function of the microwave power sent to the cavity. With increasing microwave power, the optical pump mode depletes and the conversion efficiency decreases. A linear fit in the undepleted regime (grey) yields a slope of $P_+/P_\Omega = 23.68 \pm 0.46$ corresponding to a photon number conversion efficiency of $\eta_+ = (1.09 \pm 0.02) \times 10^{-3}$. The inset shows an OSA spectrum (resolution \SI{2.5}{GHz}) of the reflected pump power and the generated upconverted sideband.}
	\label{fig:OSA}
\end{figure}

The photon number conversion efficiency can be found by measuring the microwave power at the input and the optical signal power at the output of the converter. From \eqref{power_conversion} we find:
\begin{equation}
\eta_+ = \frac{\Omega}{\omega_+} \frac{P_+}{P_\Omega}
\label{equ:photon_conversion_measurement}
\end{equation}
As input power of the microwave we take the power arriving at the coaxial pin coupler while cable losses are calibrated out. As optical signal power we consider the power leaving the WGM resonator inside the silicon prism, see Fig.~\ref{fig:setup} c). This is justified for estimating the performance of the system because the prism can be coated to prevent reflection loss.
The power of the sideband is obtained by measuring its relative strength compared with the reflected pump power on the optical spectrum analyzer. This allows us to calculate the sideband power leaving the resonator from the coupling efficiency of the pump and the absolute pump power sent to the resonator.

For demonstration of the absolute conversion efficiency, we choose the optical mode depicted in Fig.~\ref{fig:setup}(d) {and pump it on resonance}. The working temperature is set such, that the avoided crossing occurs on the red detuned side of the mode. This ensures that we can obtain a single sum-frequency generated sideband which is not disturbed by the linear mode coupling causing the avoided crossing. We measure $\text{FSR}_+ = \SI{8.941}{GHz}$ and adjust the microwave resonance frequency $\Omega_0$ to that value with the tuning screw. In this configuration we find for the undercoupled microwave mode $\gamma_\Omega = 2 \pi \times \SI{3.6}{MHz}$ and $\gamma'_\Omega = 2 \pi \times \SI{16.2}{MHz}$ and for the critically coupled optical mode $\gamma = \gamma' = 2\pi \times \SI{346}{kHz}$. The laser is locked to the resonator and \SI{1}{mW} optical power is sent into the cavity. Considering the imperfect coupling and the resulting reflection from the prism (see Fig.~\ref{fig:setup}(d)), \SI{0.42}{mW} are actually coupled into the WGM resonator. The microwave signal frequency is set on resonance and increased from \SI{-54}{dBm} to \SI{-22}{dBm} in steps of \SI{1}{dBm}.\\
The results are presented in Fig.~\ref{fig:OSA}. The inset shows a typical OSA spectrum, where the single up-converted optical sideband is highlighted in blue. The plot shows the optical signal power leaving the resonator as a function of the input microwave signal power. In the undepleted pump regime, the optical signal scales linearly with the input microwave power according to \eqref{power_conversion}. At approximately \SI{1}{\uW} microwave power, the pump starts to deplete and the up-converted signal saturates. From fitting the linear regime we find a power conversion efficiency of $P_+/P_\Omega = 23.68 \pm 0.46$ which transforms into an absolute photon number conversion efficiency of 
$\eta_+ = (1.09 \pm 0.02) \times 10^{-3}$ using \eqref{equ:photon_conversion_measurement}. {Inserting the cavity parameters into \eqref{power_conversion},} we find the single photon coupling rate to be $g\nii{eff} = 2\pi\times (7.43 \pm 0.02)$\,Hz. 
In order to compare this value with the theoretically derived one, we have to multiply it with $\sqrt{2}$ since the theory does not take into account that we experimentally excite propagating and counterpropagating modes at the same time.\\
 This effective coupling rate is about three times smaller than the expected one from numerical simulations, corresponding to almost ten times decreased conversion efficiency. This is {likely} caused by air gaps between the WGM resonator and the metallic rings clamping the resonator. Due to the high dielectric constant of lithium niobate in the microwave regime even very small air gaps lead to a significant field drop in air. Our simulations show that an air gap of \SI{20}{\um} caused by imperfect machining of the cavity and polishing of the upper and lower side of the WGM resonator lowers $g$ already by a factor of two.

\section{Discussion}
The three main results of this paper are the following. The first result is a three order of magnitude improvement of resonant electro-optic conversion of microwave photons into the optical regime compared to the so far best reported values \cite{ilchenko_whispering-gallery-mode_2003,strekalov_microwave_2009}. The second result is a highly efficient suppression of either the up- or downconverted sideband by engineering the dispersion of the resonator. {The third result is a MHz bandwidth for the conversion. All three} are important steps towards coherent {and bidirectional} microwave photon conversion based on direct electro-optic modulation. {For quantum operations, the conversion efficiency will have to be pushed towards unity, which is possible with some engineering improvements.}
Tsang \cite{tsang_cavity_2011} introduces the electro-optic cooperativity $G_0 = |g \alpha|^2/(\Gamma_+ \Gamma_\Omega)$, where $|\alpha|^2$ is the number of pump photons within the resonator. For our system we find $G_0 \approx 4\times10^{-3}$, which is almost three orders of magnitude smaller than required for a device suitable for converting non-classical photon states requiring $G_0 \approx 1$. To enhance the cooperativity the pump power can be increased only slightly, as photo- and thermorefractive effects in lithium niobate cause the system to become unstable. While the optical $Q$ factors are already close to the material limitation \cite{leidinger_comparative_2015}, it was shown recently that the intrinsic microwave absorption in lithium niobate at millikelvin temperatures allows for $Q_\Omega \approx 10^{5}$ \cite{goryachev_single-photon_2015}. This alone would increase our $G_0$ by three orders of magnitude. 
Furthermore, simulations show, that $g$ scales inversely with the resonator thickness. Assuming a realistic resonator thickness of \SI{50}{\um}, together with closing any air gaps deteriorating $g$, {can improve the cooperativity additionally by two to three orders of magnitude}. Recently, a theoretical study showed further routes to increase $g$ \cite{javerzac-galy_-chip_2015}. Putting these optimizations together, $G_0 \gg 1$ {can be achieved in a realistic scenario.} \\
{Electro-optomechanical schemes work in the resolved sideband regime where two pump tones, an optical and a microwave one, are detuned from their respective resonance frequencies by the mechanical resonance frequency. The consequence is that the bandwidth of the process is ultimately limited to the resonance frequency of the mechanical resonator which is typically below \SI{1}{MHz}. In practice, only tens of kilohertz have so far been achieved in efficient conversion experiments \cite{andrews_bidirectional_2014}. Coupling such systems to on demand single microwave photon sources \cite{bozyigit_antibunching_2011} is challenging as the temporal signature of these photons is in the order of MHz \cite{andrews_quantum-enabled_2015}.}
Furthermore, the achievable suppression of the red-detuned sideband is limited as the pump detuning is also determined by the mechanical resonance frequency. Depending on the system parameters this can lead to contamination of the converted signal with spontaneous emission.\\
{Our system does not have these restrictions as the mediator between the two regimes is the non-resonant, nonlinear polarizability of lithium niobate. Hence, the bandwidth of the electro-optic scheme is solely determined by intrinsic losses and the nonlinear as well as external coupling rates.
Furthermore, the described dispersion management of the optical modes allows us to choose the degree of suppression freely without detuning the optical pump.}\\
These advantages justify the further investigation and optimization of the electro-optical approach, even though the electro-optomechanical presently has the efficiency advantage.
As we have discussed, the single photon regime is within reach provided an optimized system. This would not only allow interfacing microwave qubits with the optical domain, but also electro-optic cooling of the microwave mode. With even lesser requirements, the generation of entangled pairs of microwave and optical photons utilizing parametric down conversion is possible, which has recently been shown in the optical domain in a WGM based system \cite{schunk_interfacing_2015}.


\section*{Acknowledgments}
We would like to acknowledge stimulating discussions with Konrad Lehnert and Alessandro Pitanti. MCC would like to acknowledge support of the Studienstiftung des deutschen Volkes.

\end{document}